\documentstyle[onecolumn,epsfig]{mn}
\begin{document}

\title[Gravitational collapse]{Gravitational collapse of polytropic, magnetized, filamentary clouds }

\author[Mohsen Shadmehri]
{Mohsen Shadmehri\thanks{E-mail:
mshadmehri@science1.um.ac.ir}\\
Department of Physics, School of Science, Ferdowsi University,
Mashhad, Iran}

\maketitle

\date{Received 20 August 2002 / Accepted _________________ }
%
%
\markboth{Shadmehri: Gravitational collapse}{}
\begin{abstract}
When the gas of a magnetized filamentary cloud obeys a polytropic
equation of state, gravitational collapse of the cloud is studied
using a simplified model. We concentrate on the radial
distribution and restrict ourselves to the purely toroidal
magnetic field. If the axial motions and poloidal magnetic fields
are sufficiently weak, we could reasonably expect our solutions
to be a good approximation. We show that while the filament
experiences gravitational condensation and  the density at the
center increases, the toroidal flux-to-mass ratio remains
constant. A series of spatial profiles of density, velocity and
magnetic field for several values of the toroidal flux-to-mass
ratio and the polytropic index, is obtained numerically and
discussed.
\end{abstract}

%

\section{Introduction}

   Understanding the processes of transforming molecular clouds into
stars is one of the main goals of the people who are working on
the structure formation in interstellar medium (ISM). Since
various physical agents such as self-gravity, thermal processes,
magnetic fields are playing significant roles in star formation,
we are still far from a coherent and consistent picture in spite
of great achievements  during recent years. While the clouds in
the standard model of star formation  are considered to be
initially in equilibrium or quasi-hydrostatic phase, some authors
questions whether this important aspect is  plausible according
to the numerical simulations and the observations (e.g,
V\'{a}zquez-Semadeni at al. 2004). Irrespective of which theory
can correctly describe the initial state of the clouds, the next
stage of evolution of the cloud is gravitational collapse. There
are many studies for presenting a correct description of the
gravitational collapse of the clouds, considering their
geometrical shapes and the main physical factors.

Studies of the gravitational collapse of gaseous clouds have been
started by the pioneer works of Bodenheimer $\&$ Sweigart (1968),
Larson (1969), and Penston (1969) who studied the isothermal
collapse of spherical clouds using numerical integration of the
equations and semi-analytical similarity solutions. It seems that
self-similar flows provide the basic physical insights to the
gravitational collapse, and may indicate the way to more detailed
investigations (Shu 1977; Hunter 1977, 1986; Whitworth $\&$
Summers 1985).

Most of the previous works dedicated to study of the collapse of
spherically symmetric clouds. But it is known that filamentary
structures associated with clumps and cores are very common
(e.g., Schneider \&\ Elmegreen 1979; Houlahan \&\ Scalo 1992;
Harjunpaa et al. 1999). Also, filamentary structures are so
prominent in numerical simulations of star formation that seems
the formation and fragmentation of filaments is an important
stage of star formation (see, e.g., Jappsen at al. 2004). So,
self-similar solutions for the collapse of a filamentary cloud
were investigated, and different sets have been found (Inutsuka
\&\ Miyama 1992; Kawachi \&\ Hanawa 1998; Semelin, Sanchez, \&\ de
Vega 1999; Shadmehri \&\ Ghanbari 2001; Hennebelle 2003; Tilley
\&\ Pudritz 2003 ). Recently, Hennebelle (2003) investigated
self-similar collapse of an isothermal magnetized filamentary
cloud which may undergo collapse in axial direction, in addition
to radial direction. Then, Tilley \&\ Pudritz (2003) extended
this analysis by focusing on only the purely radial motions and
obtained interesting analytical solutions able to describe
collapse of an isothermal magnetized filamentary cloud with
purely toroidal magnetic field.

While most of the authors assumed that the clouds are isothermal
and so their self-similar solutions describe collapse of
isothermal filamentary clouds, Shadmehri \&\ Ghanbari (2001)
studied quasi-hydrostatic cooling flows in filamentary clouds
using similarity method. Although an isothermal equation of state
is a natural first approximation, there are some growing
evidences that precise isothermality is not expected in molecular
clouds. Scalo et al. (1998) studied the likely values of the
exponent $\gamma$ that appears in polytropic equation of state of
the form $P=K \rho^{\gamma}$. Kawachi \&\ Hanawa (1998)
investigated the gravitational collapse of a nonmagnetized
filamentary cloud using zooming coordinates (Bouquet et al.
1985). They used a polytropic equation of state to indicate the
effects of deviations from isothermality in the collapse.

In this paper we derive self-similar solutions that can describe
collapse of a polytropic filamentary cloud considering magnetic
effects. While in the isothermal case the radial velocity is in
proportion to the radial distance and one has to choose a
prescription between magnetic field and the density (Tilley \&\
Pudritz 2003; hereafter TP), we show that when the gas obeys a
polytropic relation not only the radial velocity has not such
simple nature but also the toroidal flux-to-mass ratio is
constant during the collapse phase. The equations of the model
are presented in the second section. We obtain and solve the set
of self-similar equations in the third section. These solutions
will be discussed in this section.


\section{General Formulation}

%
   In order to study gravitational collapse of filamentary
magnetized clouds, we start by writing the equations of ideal
magnetohydrodynamics in cylindrical coordinates $(r, \varphi,
z)$. We consider axisymmetric and long filament along the $z$
axis. Thus, all the physical variables depend just on the radial
distance $r$ and time $t$. As for the magnetic field geometry,
the toroidal component $B_{\rm\varphi}$ of the field is assumed to
be dominant. The governing equations are the continuity,
\begin{equation}
\frac{\partial\rho}{\partial t} +
\frac{1}{r}\frac{\partial}{\partial r}(r \rho v_{\rm r})=0,
\end{equation}
the momentum equation,
\begin{displaymath}
\frac{\partial v_{\rm r}}{\partial t} + v_{\rm r}\frac{\partial
v_{\rm r} }{\partial r} +\frac{1}{\rho}\frac{\partial p}{\partial
r } + \frac{\partial \Psi}{\partial r}=-
\frac{B_{\rm\varphi}}{\mu \rho r}\frac{\partial}{\partial r}(r
B_{\rm\varphi}),
\end{displaymath}
the Poisson's equation,
\begin{equation}
\frac{1}{r}\frac{\partial}{\partial r}(r
\frac{\partial\Psi}{\partial r})=4\pi G\rho,
\end{equation}
and the induction equation,
\begin{equation}
\frac{\partial B_{\rm\varphi}}{\partial
t}+\frac{\partial}{\partial r}(v_{\rm r}B_{\rm\varphi})=0,
\end{equation}
We also assume a polytropic relation between the gas pressure and
the density,
\begin{equation}
p= K \rho^{\gamma},
\end{equation}
with $K$ and $\gamma=1+\frac{1}{n}$ are constants and $\gamma <
1$ (Scalo et al. 1998). So, we relax the isothermal approximation
adopted in the previous study by TP. Since we neglected the axial
velocity $v_{\rm z}$, the corresponding solutions could only be
applied to the regions near the middle of a filament having
finite length.
\section{Self-similar solutions}
\subsection{similarity equations}
In the self-similar formulation, the various physical quantities
are expressed as dimensionless functions of a similarity
variable. The two-dimensional parameter of the problem are $K$
and Newton's constant, $G$, from which we can construct a unique
similarity varaible
\begin{equation}
\xi=K^{-\frac{1}{2}} G^{\frac{\gamma-1}{2}} r
(t_0-t)^{\gamma-2},{\label{eq:sim}}
\end{equation}
where $t<t_0$ and the term, $t_{0}$, denotes an epoch at which the
central density increases infinitely. Dimensionless density,
velocity, gravitational potential, and toroidal component of
magnetic field then be set up as
\begin{equation}
\rho (r, t) = G^{-1} (t_0-t)^{-2} R(\xi),{\label{eq:density}}
\end{equation}
\begin{equation}
p (r, t) =K G^{-\gamma} (t_0-t)^{-2\gamma} P(\xi),
\end{equation}
\begin{equation}
\ v_{\rm r}(r, t) = K^{\frac{1}{2}} G^{\frac{1-\gamma}{2}}
(t_0-t)^{1-\gamma}V(\xi),
\end{equation}
\begin{equation}
\ B_{\rm\varphi}(r, t) = \mu^{\frac{1}{2}} K^{\frac{1}{2}}
G^{-\frac{\gamma}{2}}  (t_0-t)^{-\gamma}
b_{\rm\varphi}(\xi),{\label{eq:Bphi}}
\end{equation}
\begin{equation}
\Psi(r, t) = K G^{1-\gamma} (t_0-t)^{2(1-\gamma)} S(\xi).
\end{equation}
Thus, in terms of these similarity functions, the equation of
state simply becomes $P=R^{\gamma}$. The continuity equation,
Euler equation, Poisson equation, and induction  equation become
\begin{equation}
2R + (2-\gamma)\xi \frac{dR}{d\xi}+\frac{1}{\xi}\frac{d}{d\xi}(\xi
RV)=0,{\label{eq:con}}
\end{equation}
\begin{displaymath}
(\gamma-1)V + (2-\gamma)\xi
\frac{dV}{d\xi}+V\frac{dV}{d\xi}+\frac{1}{R}\frac{dP}{d\xi}
\end{displaymath}
\begin{equation}
+\frac{dS}{d\xi}=-\frac{b_{\rm\varphi}}{\xi R}\frac{d}{d\xi}(\xi
b_{\rm\varphi}),\label{eq:motion}
\end{equation}
\begin{equation}
\frac{1}{\xi}\frac{d}{d\xi}(\xi\frac{dS}{d\xi})=4\pi
R,\label{eq:possion}
\end{equation}
\begin{equation}
\gamma
b_{\rm\varphi}+(2-\gamma)\xi\frac{db_{\rm\varphi}}{d\xi}+\frac{d}{d\xi}(V
b_{\rm\varphi})=0,{\label{eq:bp}}
\end{equation}

We can obtain solutions of TP for the collapse of an isothermal
magnetized filament, simply by substituting $\gamma=1$ in the
above equations. In this case, equation (\ref{eq:con}) is
integrable and gives $V=-\xi$. Thus, equation (\ref{eq:bp}) is
automatically satisfied and one has to choose a relationship
between the magnetic field and the density in order to make
further progress (TP). Clearly, in the polytropic case $\gamma
\neq 1$, the continuity equation (\ref{eq:con}) has not the
simple nature of the isothermal collapse. However, from Equations
(\ref{eq:con}) and (\ref{eq:bp}) one can simply show that the
toroidal component of magnetic field, $b_{\rm\varphi}$, should be
proportional to $\xi R$. On the other hand, the toroidal
flux-to-mass ratio $\Gamma_{\rm\varphi}$ is defined as
$\Gamma_{\varphi}=B_{\rm\varphi}/r\rho$ (Fiege \& Pudritz 2000).
Considering equations (\ref{eq:sim}), (\ref{eq:density}) and
(\ref{eq:Bphi}), we can write $\Gamma_{\rm\varphi}$ as
\begin{equation}
\Gamma_{\varphi}=\frac{b_{\rm\varphi}}{\xi R}.\label{eq:gamp}
\end{equation}
Since the above similarity equations show   that $b_{\rm\varphi}$
is in proportion to $\xi R$, the toroidal flux-to-mass ratio
$\Gamma_{\rm\varphi}$ should be constant. We can consider
$\Gamma_{\rm\varphi}$ as free parameter.

If we write
\begin{equation}
U=(2-\gamma)\xi+V,
\end{equation}
then we can re-write the equations as
\begin{equation}
\frac{dR}{d\xi}=\frac{RX}{D},\label{eq:m1}
\end{equation}
\begin{equation}
\frac{dV}{d\xi}=-\frac{(2+V/\xi)D+UX}{D},\label{eq:m2}
\end{equation}
where
\begin{equation}
D=\gamma R^{\gamma-1}-U^{2}+\Gamma_{\varphi}^{2}\xi^{2}R,
\end{equation}
\begin{equation}
X=(2+\frac{V}{\xi})U+(1-\gamma)V-\frac{2\pi}{1-\gamma}RU-2\Gamma_{\varphi}^{2}\xi
R.
\end{equation}
Equations (\ref{eq:gamp}), (\ref{eq:m1}) and (\ref{eq:m2})
describe the gravitational collapse of a polytropic magnetized
filament. These equations are very similar to the equations
obtained by Larson and Penston for the collapse of a spherical
non-rotating and non-magnetized cloud (Larson 1969; Penston 1969)
and  extensively investigated by Hunter (1977), Shu (1977) and
Whitworth \& Summers (1985). If we set $\Gamma_{\rm\varphi}=0$,
the equations reduce to the equations derived by Kawachi and
Hanawa (1998) for the collapse of a polytropic unmagnetized
filament. The aim of this paper is to solve these coupled
differential equations (\ref{eq:m1}) and (\ref{eq:m2}) subject to
suitable boundary condition, and interpret the solutions. When
$D$ vanishes, all the numerators in equations (\ref{eq:m1}) and
(\ref{eq:m2}) must vanish at the same time; otherwise the
equations become singular and will yield unphysical solutions.

\subsection{special analytical and asymptotic solutions}
Equations (\ref{eq:m1}) and (\ref{eq:m2}) have an analytical
solution,
\begin{equation}
V=0,
\end{equation}
\begin{equation}
R=A \xi^{\frac{-2}{2-\gamma}},
\end{equation}
\begin{equation}
b_{\rm\varphi}=\Gamma_{\rm\varphi}A
\xi^{-\frac{\gamma}{2-\gamma}},
\end{equation}
where $A$ is a constant. In dimensional units, this solution
corresponds to time-independent singular configuration.

\begin{figure}
\epsfig{figure=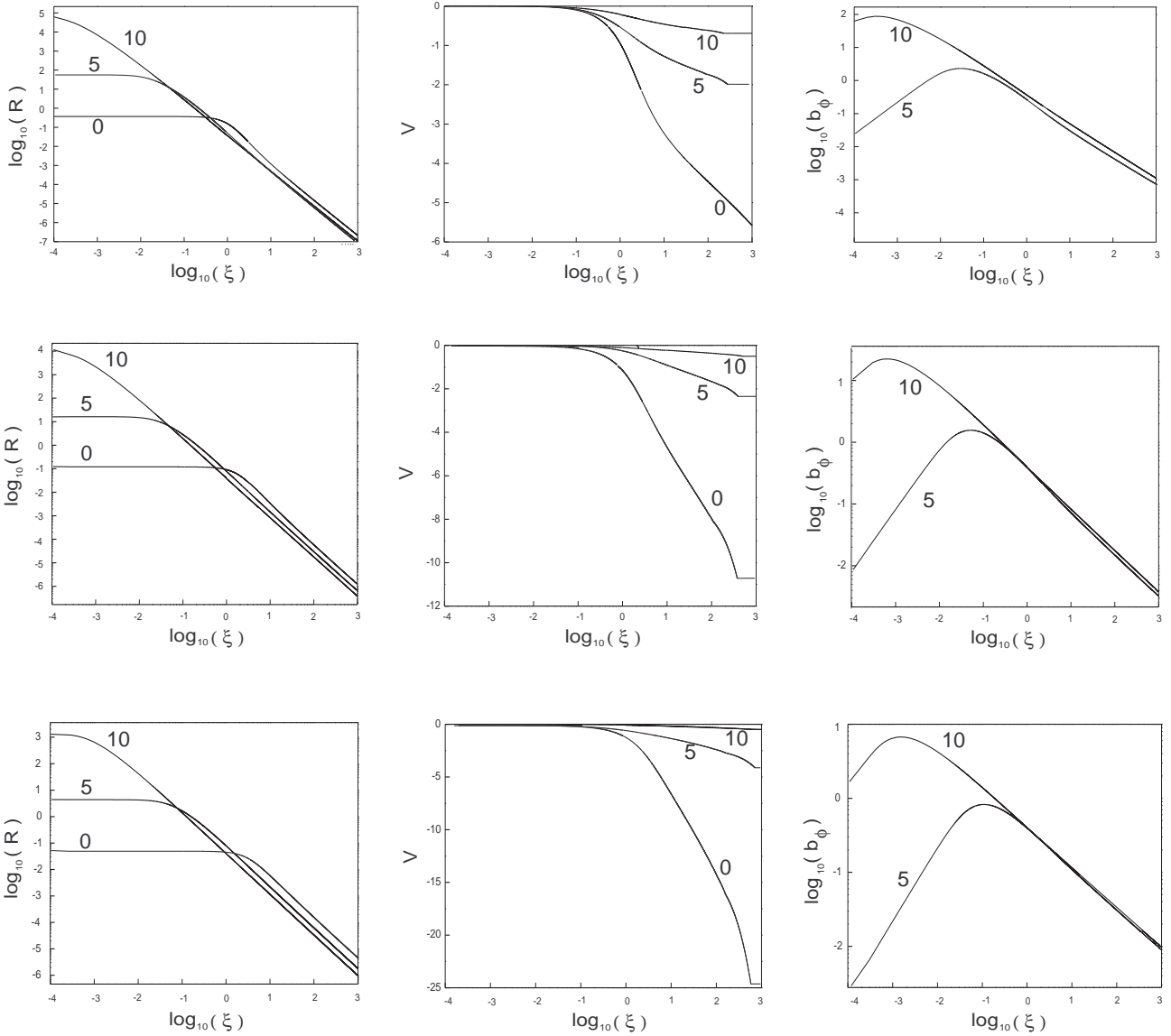,angle=0,width=\hsize} \caption{Profiles of
the density, the radial velocity and the toroidal component of
the magnetic field corresponding to $\gamma=0.9$ ({\it top}),
$0.8$ ({\it middle}) and $0.7$ ({\it bottom}). Each curve is
marked by  the toroidal flux-to-mass ratio $\Gamma_{\rm\varphi}$.
}\label{fig:figure1}
\end{figure}

We can find a family of limiting solutions for $\xi \rightarrow
0$. In order to obtain these, one can expand $R(\xi)$ and $V(\xi)$
in Taylor series
\begin{equation}
R(\xi)=R_0+R_1 \xi+R_2 \xi^2 + \ldots
\end{equation}
\begin{equation}
V(\xi)=V_0+V_1 \xi+V_2 \xi^2 + \ldots
\end{equation}
By substituting the above series in equations
(\ref{eq:con})-(\ref{eq:bp}), we obtain the following limiting
solutions for $\xi \rightarrow 0$ parameterized by $R_0$ and
$\Gamma_{\rm\varphi}$:
\begin{equation}
R(\xi)=R_0-\frac{n(\pi+\Gamma_{\rm\varphi}^2)}{n+1}
R_0^{2-\frac{1}{n}} \xi^2 + O(\xi^4),\label{eq:exp1}
\end{equation}
\begin{equation}
V(\xi)=-\xi- \frac{\pi+\Gamma_{\rm\varphi}^2}{2(n+1)}
R_0^{1-\frac{1}{n}} \xi^3 + O(\xi^5),\label{eq:exp2}
\end{equation}
\begin{equation}
b_{\rm\varphi}=\Gamma_{\rm\varphi}\xi
[R_0-\frac{n(\pi+\Gamma_{\rm\varphi}^2)}{n+1} R_0^{2-\frac{1}{n}}
\xi^2 + O(\xi^4)].\label{eq:exp3}
\end{equation}
Asymptotic solutions (\ref{eq:exp1})-(\ref{eq:exp3}) are very
useful when performing numerical integrations to obtain
similarity solutions starting from $\xi \rightarrow 0$.
\begin{table*}
 \centering
 \begin{minipage}{140mm}
  \caption{Summary of the similarity solution for various $\gamma$ and $\Gamma_{\rm\varphi}$.}
  \begin{tabular}{@{}lccccccccccc@{}}
  \hline
  &&{$\gamma=0.9$}& &  & &$\gamma=0.8$&&&&$\gamma=0.7$\\
  \hline
             $\Gamma_{\rm\varphi}$&$\xi_{\ast}$ & $R_{\ast} (\times 10^{2})$ & $R_{\rm 0}$ &
            &  $\xi_{\ast}$ & $R_{\ast}(\times 10^{2})$  & $R_{\rm 0}$ &
     &  $\xi_{\ast}$ & $R_{\ast}(\times 10^{2})$  & $R_{\rm 0}$ \\

 \hline
 0 & 2.99 & $1.76$ & 0.37  &    & 3.20 & 2.57 & 0.12 &   & 3.54   & 2.29  & 0.04\\
 1 & 2.87 & $1.73$ & 0.56  &    & 3.09 & 2.54 & 0.17 &   & 3.22   & 2.62  & 0.07\\
 2 & 2.67 & $1.64$ & 1.45  &    & 2.66 & 2.58 & 0.45 &   & 2.57   & 3.17  & 0.25\\
 3 & 2.51 & $1.51$ & 4.73  &    & 2.44 & 2.40 & 1.52 &   & 2.48   & 2.82  & 0.54\\
 4 & 2.40 & $1.37$ & 16.8  &    & 2.31 & 2.16 & 4.89 &   & 2.32   & 2.56  & 1.49\\
 5 & 2.35 & $1.23$ & 63    &    & 2.24 & 1.90 & 16.32&   & 2.23   & 2.27  & 4.36 \\
 6 & 2.32 & $1.10$ & 246   &    & 2.22 & 1.66 & 56.2 &   & 2.20   & 1.98  & 12.85\\
 7 & 2.33 & $0.97$ & 980  &     & 2.23 & 1.44 & 200  &   & 2.21   & 1.70  & 39.30\\
 8 & 2.34 & $0.88$ & 3968  &    & 2.26 & 1.24 & 731  &   & 2.24   & 1.46  & 124 \\
 9 & 2.37 & $0.78$ & 17866 &    & 2.31 & 1.07 & 2798 &   & 2.29   & 1.25  & 404 \\
 10& 2.41 & $0.69$ & 62700 &    & 2.36 & 0.93 & 10446&   & 2.35   & 1.07  & 1386\\
\hline

\hline

\end{tabular}
\end{minipage}
\end{table*}

\subsection{numerical solutions}
When similarity solutions cross the critical points, we must take
extra care as for their analyticity across the critical points.
In order to find the behaviour of solutions around critical
point, $\xi=\xi_{\ast}$, we expand $R(\xi)$ and $V(\xi)$ in a
Taylor series up to first order. Detailed calculations are in
Appendix A. Given $\gamma$ and $\Gamma_{\rm\varphi}$, we can find
analytical expressions for $\xi_{\ast}$ and $V_{\ast}$ as
functions of $R_{\ast}$. However, note that only those points are
acceptable that $\xi_{\ast}>0$ and $V_{\ast}<0$. Although we
don't study type of critical points in detail, when performing
numerical integrations we note if the critical points are saddle
or nodal  (Jordan \& Smith 1977). In general, the critical point
is a saddle point when the two gradients of eigensolutions of
$V(\xi)$ are of the opposite signs. On the other hand, when the
two gradients have the same sign, the critical point is a nodal
point.

We can numerically integrate equations (\ref{eq:m1}) and
(\ref{eq:m2}) by a fourth order Runge-Kutta integrator. By using
asymptotic solutions (\ref{eq:exp1}) and (\ref{eq:exp2}) for $\xi
\rightarrow 0$, one can start  integration of equations
(\ref{eq:m1}) and (\ref{eq:m2}) at a very small $\xi$ ( e.g.,
$\xi=10^{-4}$) and with arbitrary value for $R_{0}$ for given
$\Gamma_{\rm\varphi}$ and $\gamma$. Also, we can integrate the
same equations (\ref{eq:m1}) and (\ref{eq:m2}) backward from the
critical position, using asymptotic behaviour of solutions near
the critical point for an arbitrary $R_{\ast}$. Thus, we look for
matchings of $R$ and $V$ at some specific point $\xi_{\rm m}$
($0<\xi_{\rm m} < \xi_{\ast}$). This can be found by studying
loci in $V-R$ plane at $\xi_{\rm m}$.

Table 1 summarizes the dependence of the similarity solutions on
$\Gamma_{\rm\varphi}$ and $\gamma$. Listed are  the position of
the critical point ($\xi_{\ast}$), the density at this point
($R_{\ast}$) and the density at the center ($R_{0}$). Figure
\ref{fig:figure1} shows profiles of the  density $R$, the radial
velocity $V$ and the toroidal component of the magnetic field
$b_{\rm\varphi}$ for different choice of $\gamma$ and
$\Gamma_{\rm\varphi}$. Clearly the solutions except for the
geometry and the magnetic field are equivalent to the
Larson-Penston solution. In almost all cases, there are an inner
part with approximately flat density profile and an outer region
with decreasing density. Since the toroidal flux-to-mass ratio
$\Gamma_{\rm\varphi}$ is constant during the collapse, each curve
in Figure \ref{fig:figure1} is labeled by this ratio. For a fixed
$\Gamma_{\rm\varphi}$, the central density increases as $\gamma$
increases. The infall velocity increases and tends to an
asymptotic value at large radii. Of course, such large velocities
are not acceptable and there must be a mechanism (e.g., pressure
turncation) to turncate the solutions at finite radius. However,
the typical behaviour of the radial velocity of a collapsing
polytropic filament is different from the isothermal collapse
where the self-similar infall velocity behaves in proportion to
the radial distance. Also, Figure \ref{fig:figure1} shows that in
the inner part the toroidal component of the magnetic field
$b_{\rm\varphi}$ increases, while in the outer part
$b_{\rm\varphi}$ is a decreasing function of $\xi$. This
behaviour is easily understood, if we note $b_{\rm\varphi}
=\Gamma_{\rm\varphi} \xi R$. For example, in the inner part, the
density is roughly constant and so $b_{\rm\varphi} \propto\xi$.

An interesting feature of the solutions is that the cloud is more
compressed by the toroidal pinching, if the toroidal magnetic
field increases. Figure \ref{fig:figure1} shows that the size of
the inner region strongly depends on $\Gamma_{\rm\varphi}$.
However, the typical behaviours of the density and the magnetic
field in the outer part are more or less independent of the
toroidal field. We see that the central density increases, if
$\Gamma_{\rm\varphi}$ increases so that in highly magnetized
filament, the inner part with flat density profile disappears and
joins to the outer region. If we compare the density profiles of
$\gamma=0.9$ and $\gamma=0.7$ for $\Gamma_{\rm\varphi}=10$, we see
there is still a small inner region for lower  $\gamma=0.7$. It
simply implies as $\gamma$ decreases, there needs higher level of
magnetic intensity in order to affect the flat density profile of
the inner part. In other words, as the cloud tends to the
isothermal regime, the inner part becomes more sensitive on the
toroidal magnetic field.

The infall velocity reduces as $\Gamma_{\rm\varphi}$ increases.
While in nonmagnetic collapse, the radial velocity tends to very
high value at large radii, we see that the velocity significantly
reduces even at large radii in highly magnetized collapsing
filament. However, the density profile in the outer region hardly
depends on $\Gamma_{\rm\varphi}$. We find this profile mostly
depends on $\gamma$, so that it behaves in proportion to
$r^{-1.81}$, $r^{-1.69}$ and $r^{-1.55}$ at large radii  for
$\gamma=0.9$, $0.8$ and $0.7$, respectively. This typical
behaviour of the density is in good agreement with the asymptotic
solution at large radii which is expressed as
$R(\xi)=R_{\infty}\xi^{-2/(2-\gamma)}$. Behaviour of the density
profile at large radii does not change by increasing
$\Gamma_{\rm\varphi}$ for a given $\gamma$. This behaviour is
different from previous studies. Stod\'{o}lkiewicz (1963) and
Ostriker (1964) studied isothermal equilibrium structure of an
isothermal filament, which the density falls off as $r^{-4}$ at
large radii. Miyama et al. (1987) derived a set of self-similar
solutions for an unmagnetized collapsing filament and their
density structure is similar to unmagnetized equilibrium
solutions, i.e. in proportion to $r^{-4}$ at large radii. TP
showed that the density profile of an isothermal collapsing
filament with purely toroidal magnetic field may change at large
radii depending on the level of magnetization from $r^{-4}$ (low
magnetization) to $r^{-2}$ (high magnetization). But we see as
the cloud deviates from isothermality, these typical behaviours
of the density profile at large radii change and the main
parameter in shaping the profile is $\gamma$ not the level of
magnetization, i.e. $\Gamma_{\rm\varphi}$.

Considering profile of the density in outer part of the filament,
we can obtain the toroidal component of the magnetic field. Since
at large radii we have $R\propto r^{\nu}$, the toroidal field
becomes $b_{\rm\varphi}\propto r^{\nu+1}$, where $\nu=-1.81$,
$-1.69$ and $-1.55$  for $\gamma=0.9$, $0.8$ and $0.7$,
respectively. We discussed that the density profile at large
radii is fairly insensitive to $\Gamma_{\rm\varphi}$, and it
simply implies that the scaling of $b_{\rm\varphi}$ with the
radial distance in the outer part is also insensitive to
$\Gamma_{\rm\varphi}$. We can find behaviour of the ratio of the
thermal to the magnetic pressures, $\beta$. One can easily show
that in our notation the ratio is
$\beta=(2/\Gamma_{\rm\varphi}^2)(R^{\gamma-2}/\xi^2)$. Thus, we
see that in the inner part $\beta \propto r^{-2}$, irrespective of
the value of $\gamma$ and $\Gamma_{\rm\varphi}$. But in the outer
region we have $\beta \propto r^{(\gamma-2)\nu-2}$. These
behaviours of $\beta$ show that during gravitational collapse of a
polytropic, magnetized filament this ratio is not constant.
However, the toroidal flux-to-mass ratio is conserved during
collapse phase. In isothermal regime, one can assume either
$\beta=constant$ or $\Gamma_{\rm\varphi}=constant$ and then study
gravitational collapse (TP).
\section{conclusion}
In this paper, we have derived solutions able to describe
dynamics of a collapsing, polytropic, magnetized, self-gravitating
filament. The magnetic field is assumed to be purely  toroidal,
although it is the only possible magnetic configuration which can
be studied using the similarity method (see below).  But we expect
our solutions to be a good approximation, if the poloidal fields
are sufficiently weak. Moreover, purely toroidal field makes it
easier to compare the obtained solutions with the isothermal
collapse solutions as TP studied. Also, The self-similar
solutions of our model can be used for future numerical or
analytical studies because except for the geometry and
magnetization they resemble to the Larson-Penston solution that
has been found to be in good agreement with numerical analysis
(e.g., Larson 1969; Hunter 1977; Foster \& Chevalier 1993).

One may ask  is it possible to extend this analysis by considering
both the poloidal and the toroidal components of the magnetic
field. We note that flux conservation imposes $B_{\rm
z}\propto\rho$, which means the similarity solutions should
support this scaling. However, such similarity solutions should
be in these forms, according to a dimensional analysis: $B_{\rm z
}= \mu^{\frac{1}{2}} K^{\frac{1}{2}} G^{-\frac{\gamma}{2}}
(t_0-t)^{-\gamma} b_{\rm z}(\xi)$ and $\rho (r, t) = G^{-1}
(t_0-t)^{-2} R(\xi)$. Obviously, this scaling does not support
the flux conservation and so, it is necessary to put $b_{\rm z}=0$
or $B_{\rm z}=0$ . On the other hand, Hennebelle (2002) explored a
set of self-similar solutions for a magnetized filamentary cloud,
in which the toroidal and the poloidal components of the magnetic
field play significant role in the dynamical collapse of the
cloud. However, his solutions describe an isothermal, magnetized
filamentary cloud which undergo collapse in the axial direction,
in addition to radial collapse. In his model, the slope of the
axial velocity being two times the slope of the radial one at the
origin.

One of the major conclusions of our study is that the dynamics of
a polytropic filament is different from the isothermal case. We
showed that the toroidal component of the magnetic field help to
confine the gas by hoop stress and this conclusion is independent
of the exponent of the polytropic equation of state. Most
measurements of molecular clouds have difficulty resolving the
inner regions of filaments and so, it is very important to
understand behaviour of the physical quantities at large radii.
Our solutions showed that the typical behavior of the density of
a magnetized polytropic filament mainly depends on  polytrop
index, $\gamma$, irrespective of the level of magnetization.
However, the infall velocity in the outer regions strongly
depends on the ratio of the flux-to-mass ratio
$\Gamma_{\rm\varphi}$. It implies to examine the profiles of
molecular lines for infall velocity which may help us to
understand the true nature of the filaments. While radial
velocity of the isothermal model of TP has a very simple
behaviour, our polytropic solutions and Hennebelle (2002) model
clearly present different profiles of the radial velocity similar
to Larson-Penston solutions of collapsing unmagnetized spherical
clouds.

The environment of L1512, a starless core, has been studied at
high angular resolution by Falgarone, Pety \& Philips (2001). The
gas outside the dense core is structured in several filaments
with a broad range of density and temperature, from $n_{\rm
H_{\rm 2}}=2\times 10^{3}$ cm$^{-3}$ for the coldest case ($T=20$
K) down to $n_{\rm H_{\rm 2}}=180$ cm$^{-3}$ for the warmest
($T=250$ K). It suggests that a polytropic equation of state is
more appropriate for describing the thermal behaviour of
filaments of this system. Falgarone et al. (2001) discussed that
these filaments are not held either by the pressure of the H
$_{\rm I}$ layer or by the external pressure. So, they concluded
that the toroidal component helps confine the filaments. However,
their arguments were based on the analysis of Fiege \& Pudritz
(2000), in which the equilibria of pressure-truncated isothermal
and logatropic filaments held by self-gravity and helical
magnetic fields have been studied. However, it is unlikely that
the filaments in ISM are being truly static structures, and many
numerical simulations show dynamics structures in a star forming
region (e.g., Vazquez-Semadeni et al. 2005; Jappsen et al. 2004).
We think dynamics models, like what studied here, are more
adequate for filamentary structures such as those in the
environment of L1512 which show a wide range of density and
temperature.

Acknowledgements: I thank the referee, Anthony Whitworth, for a
careful reading of the manuscript and comments that lead to
improvement of the paper.

\appendix
\section[]{}
To obtain accurate transonic solution, it is useful to analyze
the behavior of the flow near the sonic point, $\xi_{\ast} > 0$.
The values of $R_{\ast}=R(\xi_{\ast})$ and
$V_{\ast}=V(\xi_{\ast})$, where $V_{\ast} < 0$, are completely
determined by requiring both the denominator and numerator of
equations (\ref{eq:m1}) and (\ref{eq:m2}) to vanish at
$\xi_{\ast}$:
\begin{equation}
\gamma
R_{\ast}^{\gamma-1}-U_{\ast}^{2}+\Gamma_{\rm\varphi}^{2}\xi_{\ast}^{2}R_{\ast}=0,\label{eq:crit1}
\end{equation}
\begin{equation}
(2+\frac{V_{\ast}}{\xi_{\ast}})U_{\ast}+(1-\gamma)V_{\ast}-\frac{2\pi}{1-\gamma}R_{\ast}U_{\ast}-
2\Gamma_{\rm\varphi}^{2}\xi_{\ast}R_{\ast}=0,\label{eq:crit2}
\end{equation}
where $U_{\ast}=(2-\gamma)\xi_{\ast}+V_{\ast}$. In general, given
$R_{\ast}$, it is a simple matter to find $\xi_{\ast}$ and
$V_{\ast}$ from equations (\ref{eq:crit1}) and (\ref{eq:crit2}).
After mathematical manipulation, we obtain
\begin{equation}
\xi_{\ast}=[\frac{-(ca+f)\pm\sqrt{(ca+f)^2+4h(cb-d)}}{2(cb-d)}]^{1/2}
\end{equation}
where
\begin{displaymath}
a=\gamma R_{\ast}^{\gamma-1}, b=\Gamma_{\rm\varphi}^{2}R_{\ast},
c=(1-\frac{2\pi}{1-\gamma}R_{\ast})^2,
\end{displaymath}
\begin{displaymath}
d=[(1-\gamma)(2-\gamma)+\Gamma_{\rm\varphi}^2 R_{\ast}]^2, f=2 a
d^{1/2}, h=a^2.
\end{displaymath}

Now, we set
\begin{displaymath}
R(\xi)=R_{\ast}+ a_{1}(\xi-\xi_{\ast})+ O[(\xi-\xi_{\ast})^2],
\end{displaymath}
\begin{displaymath}
V(\xi)=V_{\ast}+ b_{1}(\xi-\xi_{\ast})+ O[(\xi-\xi_{\ast})^2],
\end{displaymath}
in neighborhood of $\xi_{\ast}$. Substitution of this first-order
expansion in equations (\ref{eq:m1}) and (\ref{eq:m2}) leads to
the following equations for the slopes $a_1$ and $b_1$:
\begin{equation}
a_{1}D_{1}-R_{\ast}X_{1}=0,\label{eq:a1}
\end{equation}
\begin{equation}
(2+b_1+\frac{V_{\ast}}{\xi_{\ast}})D_1+U_{\ast}X_1=0,\label{eq:a2}
\end{equation}
where
\begin{displaymath}
X_{\rm
1}=(\frac{b_1}{\xi_{\ast}}-\frac{V_{\ast}}{\xi_{\ast}^2})U_{\ast}+(2+\frac{V_{\ast}}{\xi_{\ast}})(2-\gamma+b_1)+
(1-\gamma)b_1-
\end{displaymath}
\begin{displaymath}
\frac{2\pi}{1-\gamma}[a_{1}U_{\ast}+R_{\ast}(2-\gamma+b_1)]-2\Gamma_{\rm\varphi}^{2}(R_{\ast}+a_{1}\xi_{\ast}),
\end{displaymath}
\begin{displaymath}
D_1=\gamma(\gamma-1)R_{\ast}^{\gamma-2}a_{1}-2(2-\gamma+b_1)U_{\ast}+
\end{displaymath}
\begin{displaymath}
\Gamma_{\rm\varphi}^{2}\xi_{\ast}(a_{1}\xi_{\ast}+2R_{\ast}).
\end{displaymath}
From equation (\ref{eq:a1}), we have
\begin{equation}
a_{1}=-\frac{R_{\ast}}{U_{\ast}}(2+b_{1}+\frac{V_{\ast}}{\xi_{\ast}}),
\end{equation}
and substituting in equation (\ref{eq:a2}) gives an algebraic
equation for $b_{1}$ as
\begin{equation}
A b_{1}^2 + B b_{1} + C=0,
\end{equation}
where
\begin{displaymath}
A=-\gamma (\gamma -1) \frac{R_{\ast}^{\gamma
-1}}{U_{\ast}}-2U_{\ast}-\Gamma_{\rm\varphi}^{2} \frac{R_{\ast}
\xi_{\ast}^{2}}{U_{\ast}}, B=B_1+B_2+B_3,
C=U_{\ast}(C_1+C_2)+(2+\frac{V_{\ast}}{\xi_{\ast}})C_3,
\end{displaymath}
\begin{displaymath}
B_{1}=U_{\ast}(3-\gamma+\frac{V_{\ast}}{\xi_{\ast}}+\frac{U_{\ast}}{\xi_{\ast}}+2\Gamma_{\rm\varphi}^{2}\frac
{R_{\ast}\xi_{\ast}}{U_{\ast}}),
B_{2}=(2+\frac{V_{\ast}}{\xi_{\ast}})[A-\gamma (\gamma -1)
\frac{R_{\ast}^{\gamma -1}}{U_{\ast}}],
B_{3}=2\Gamma_{\rm\varphi}^{2}R_{\ast}\xi_{\ast}-\Gamma_{\rm\varphi}^{2}(2+\frac{V_{\ast}}{\xi_{\ast}})\frac{R_{\ast}
\xi_{\ast}^{2}}{U_{\ast}}-2(2-\gamma)U_{\ast},
\end{displaymath}
\begin{displaymath}
C_1=(2+\frac{V_{\ast}}{\xi_{\ast}})(2-\gamma+\frac{2\pi R_{\ast}
}{1-\gamma})-\frac{V_{\ast}U_{\ast}}{\xi_{\ast}^{2}},
C_2=2\Gamma_{\rm\varphi}^{2}(2+\frac{V_{\ast}}{\xi_{\ast}})\frac{R_{\ast}\xi_{\ast}}{U_{\ast}}-2[\frac{\pi
(2-\gamma) }{1-\gamma}+\Gamma_{\rm\varphi}^{2}]R_{\ast},
C_3=B_3-\gamma (\gamma -1)(2+\frac{V_{\ast}}{\xi_{\ast}})
\frac{R_{\ast}^{\gamma -1}}{U_{\ast}}.
\end{displaymath}
\end{document}